\documentclass[aps,floatfix,reprint,amssymb,english,showpacs]{revtex4-1}

\usepackage{amsmath,amssymb}
\usepackage{amsfonts,amsthm}
\usepackage{graphics}
\usepackage{graphicx}
\usepackage{hyperref}
\usepackage{color}
\usepackage{bm}
\usepackage{float}

\begin{document}
\title{Magnetic and Nonmagnetic Phases in Doped $AB_2$ $t$-$J$ Chains}

\author{R.~R.~Montenegro-Filho}
\author{M.~D.~Coutinho-Filho}
\affiliation{Laborat\'orio de F\'{\i}sica Te\'orica e Computacional, Departamento de F\'{\i}sica, Universidade Federal de Pernambuco, 50670-901, Recife-PE, Brazil}

\begin{abstract}
We discuss the rich phase diagram of doped $AB_2$ $t$-$J$ chains using data from DMRG and exact diagonalization techniques. The $J$ vs $\delta$ (hole doping) phase
diagram exhibits regions of itinerant ferrimagnetism, Incommensurate, RVB, and Nagaoka States, Phase Separation, and Luttinger Liquid (LL) Physics. Several
features are highlighted, such as the modulated ferrimagnetic structure, the occurrence of Nagaoka spin polarons in the underdoped regime and small values of
$J=4t^2/U$, where $t$ is the first-neighbor hopping amplitude and $U$ is the on-site repulsive Coulomb interaction, incommensurate structures with nonzero magnetization, 
and the strong-coupling LL physics in the high-doped regime. We also verify that relevant findings are in agreement with the corresponding
ones in the square and $n$-leg ladder lattices. In particular, we mention the instability of Nagaoka ferromagnetism against $J$ and $\delta$.
\end{abstract}
%\pacs{75.10.Pq,75.10.Jm,75.40.Mg,75.30.Kz,75.50.Gg}
\maketitle

\section{Introduction}
The $t$-$J$ version of the Hubbard Hamiltonian \cite{montorsi} is a key model for the understanding 
of strongly correlated electron systems. The model is defined through only two \textit{competing} parameters: the hopping integral $t$, which 
measures the electron delocalization through the lattice, and the exchange coupling $J=4t^2/U$, where $U>>t$ is the on-site Coulomb repulsion.
In fact, several versions of the simplest Hubbard Hamiltonian, with a single orbital at each lattice and the on-site Coulomb repulsion, 
have been extensively used to model
a variety of phenomena, such as: metal-insulator transition \cite{gebhard,*Imada}, quantum magnetism \cite{auerbach} 
and High-$T_c$ superconductivity \cite{and_sce_ht}.
Moreover, exact solutions \cite{montorsi} and rigorous results \cite{liebwu,*kor2005,RevLieb,*tasakihubrev,*revtian} have played a central role
in this endeavor.  

We emphasize Lieb's theorem \cite{PRLLIEB}, a generalization of the one by Lieb and Mattis \cite{LiebMattis} for Heisenberg systems, which asserts 
that the ground state (GS) 
total spin of a bipartite lattice at half filling
and $U>0$ is given by $S_{GS}=|N_A-N_B|/2$, where $N_A$ ($N_B$) is the number of sites on sublattice $A$ ($B$); indeed, 
Lieb's theorem has greatly enhanced the investigation of new aspects of quantum magnetism \cite{RevLieb,*tasakihubrev,*revtian}.
In particular, we mention the occurrence of \textit{ferrimagnetic} GS, in which case we select studies using Hubbard or $t$-$J$ models
\cite{PRLMDCF95,PRBTIAN,LiangWang,SierraPRB1999,*sierraprb2005, MontenegroPRB2006,OliveiraPRB2009,lopes2011,*lopes2014,*[{For Bose-Hubbard models, see }][{}] NJP,*Aoki}, 
including the Heisenberg strong-coupling limit \cite{PRLRAPOSO97,*PRBRAPOSO1999,*SW,CINV,PhysA,*YamamotoPRB2007},
on chains with $AB_2$ or $ABC$ topological structures with $S_{GS}=1/2$ per
unit cell \cite{PRLMDCF95,PRLRAPOSO97,*PRBRAPOSO1999,*SW,CINV,PRBTIAN,LiangWang,SierraPRB1999,*sierraprb2005,MontenegroPRB2006,*OliveiraPRB2009}, 
which implies ferromagnetic and antiferromagnetic long-range orders \cite{PRBTIAN}. Further, the 
inclusion of competing interactions or 
geometrical and kinetic frustration \cite{IvCONDMAT2009,frustrIn,*Nakano,*furuya,lyra1,*lyra2,*Rojas2012}, 
enlarge the classes of models, thereby allowing ground-states 
not obeying Lieb or Lieb and Mattis 
theorems. These studies have proved effective in describing magnetic and other physical properties of a variety of organic, organometallic, and inorganic 
quasi-one-dimensional compounds \cite{COUTINHO-FILHO2008,IvCONDMAT2009}. 

Of particular physical interest are doped systems, although in this case rigorous results are much rare \cite{RevLieb,*tasakihubrev,*revtian}. 
One exception is Nagaoka's theorem \cite{nagaoka}, which asserts that for $J=0$ ($U\rightarrow\infty$) the $t$-$J$ model with one hole 
added to the undoped system (half-filled band) is a \textit{fully polarized ferromagnet}, favored by the hole kinematics, if the lattice satisfies
the so-called connectivity condition \cite{tasaki}. A long-standing problem about this issue is the stability of the ferromagnetic state for finite 
hole densities and finite values of $J$. Numerical results have indicated \cite{PhysRevLett.108.126406,2D_F} that two-dimensional 
lattices display a fully polarized GS for $J=0$ and $\delta\lesssim 0.2$, where $\delta=N_h/N$, with $N_h$ ($N$) the total number 
of holes (sites); while, analytical studies \cite{Altshuler,PhysRevB.85.245113}
have suggested that this state is stable up to $J_t\sim \delta^2$.   

Further, an ubiquitous phenomenon in doped strongly correlated materials is the occurrence of inhomogeneous states, particularly 
spatial \textit{phase separation} in nano- and mesoscopic scales 
\cite{dagotto} and \textit{incommensurate} states \cite{dagotto,Inc}. In underdoped High-$T_c$ materials, dynamical and statical stripes in copper oxide planes has been the
focus of intensive research \cite{Revtranquada}. Concerning two-dimensional $t$-$J$ or Hubbard models, 
phase separation into hole-rich and no-hole regions was discussed in the large$-$ and small$-J$ limits \cite{emery}.
However the precise charge distribution in the ground state remains controversial. 
The use of distinct and refined numerical methods have pointed to striped \cite{PhysRevB.84.041108} or uniform phases \cite{PhysRevB.85.081110};  
recently, it was claimed that the origin of this issue relies on the strong competition between these phases \cite{corboz}.
For the linear $t$-$J$ Hubbard chain the physics is more clear \cite{ogatta}, and phase separation takes place for $J=2.5-3.0$, depending on the doping value,
but it is absent in the small$-J$ regime. 

In this work, we use Density Matrix Renormalization Group (DMRG) \cite{DMRG} technique and Lanczos exact diagonalization (ED) to obtain
the ground state phase diagram and the low-energy excitations properties of the doped $t$-$J$ model on $AB_2$ chains \cite{PRLMDCF95} for 
$J=0.0-0.4$. We verify the occurrence of an itinerant modulated ferrimagnetic (FERRI) phase in the underdoped regime, regions of incommensurate (IC) states and 
Nagaoka ferromagnetism (F), and two regions of phase separation (PS), in which IC and F states coexist with the resonating valence bond state (RVB), respectively.
In addition, we find that the RVB state is the stable phase at $\delta=1/3$, and identify a crossover region that ends at the onset of a Luttinger liquid
(LL) phase at $\delta=2/3$, above which the LL physics \cite{luttinger} sets in.  
\begin{figure}
\includegraphics*[width=0.4\textwidth,clip]{fig1a_rev.eps}
\includegraphics*[width=0.2\textwidth,clip]{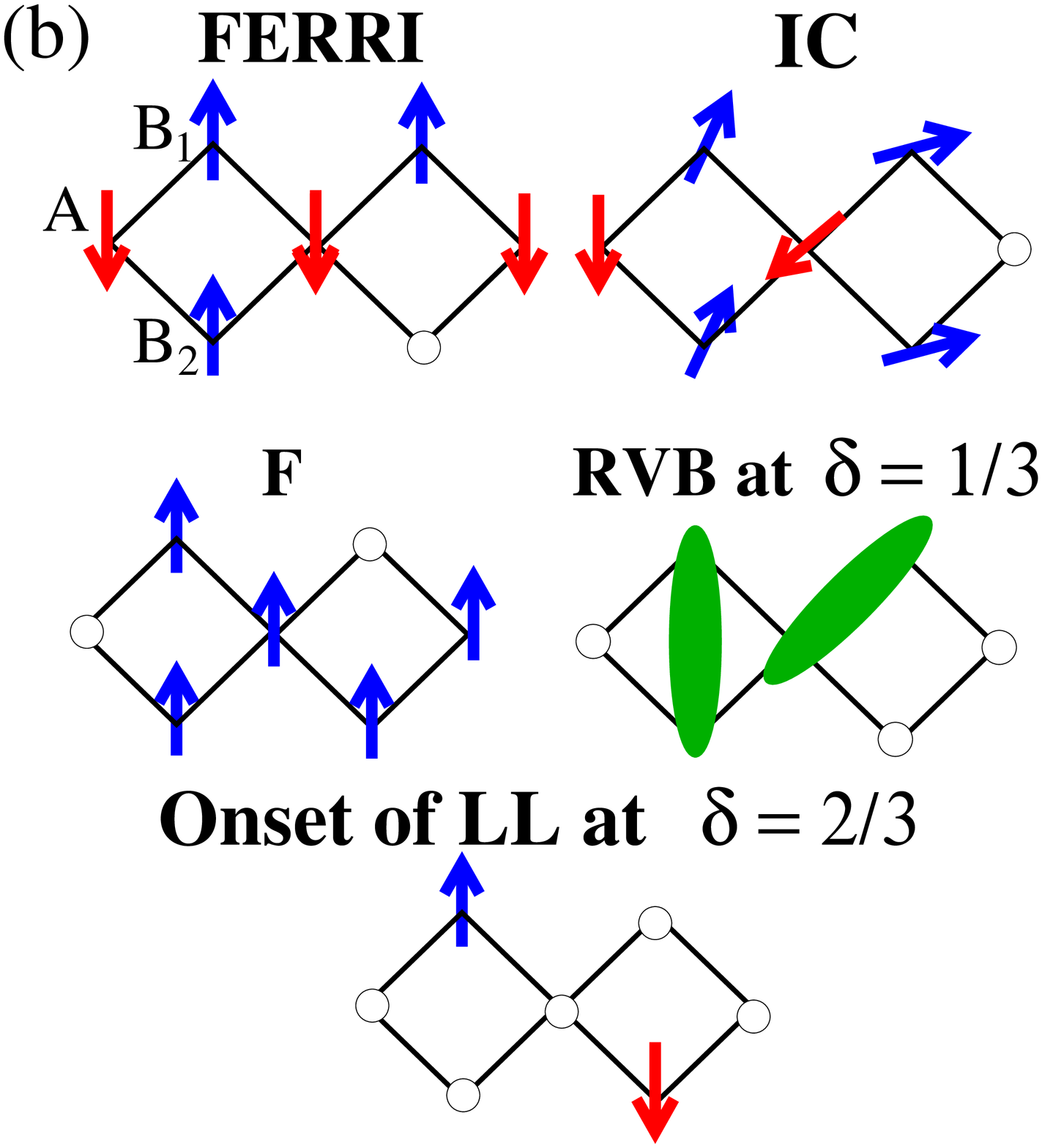}
\includegraphics*[width=0.23\textwidth,clip]{fig1c.eps}
\caption{(Color online). (a) GS phase diagram for the $AB_2$ $t$-$J$  model (error bars account for the discrete
values assumed by $\delta$ in a finite-size system). The 
phases are illustrated in (b): modulated ferrimagnetism (FERRI), incommensurate (IC), Nagaoka ferromagnetism (F),
short-range resonating valence bond (RVB) states, 
phase separation (PS), and Luttinger liquid (LL). The estimated transition lines $\delta_{FERRI,J}$, $\delta_{PS,J}$, and $J_{F,\delta}$ are also pointed out. 
(c) Ground state total spin, $S_{GS}$, normalized by its value in the undoped regime: $S_{L}\equiv(N_c/2)-0.5$, as function of $\delta$ for the indicated 
values of $J$ and $N=3N_c+1=100$.\label{diagrama}}
\end{figure}

\section{Phase Diagram}
The $t$-$J$ model reads:
\begin{align}
H_{t-J}&=-t\sum_{<i,j>,\sigma}P_G (c^{\dagger}_{i\sigma}c_{j\sigma}+H.c.)P_G\\ \nonumber
&{}+J\sum_{<i,j>}(\mathbf{S}_i\cdot\mathbf{S}_j-\frac{1}{4}n_i n_j),
\end{align}
where $c_{i\sigma}$ annihilates electrons of spin $\sigma$ at site $i$,
$n_{i}$ is the number operator at site $i$ and
$P_G=\prod_i(1-n_{i\uparrow}n_{i\downarrow})$ is the Gutzwiller projector 
operator that excludes states with doubly occupied sites. In our simulations, we set $t = 1$ and have considered chains
with $N_c$ $(N)$ unit cells (sites). In ED calculations closed boundary conditions are used with $N_c = 8$ $(N = 3N_c)$, while in the
DMRG simulations open boundary conditions are used and the system sizes ranged from $N_c = 33$ $(N = 3N_c + 1=100)$ to $N_c = 121$ $(N = 364)$. 
We retain from 243 to 364 states in the DMRG calculations, and the typical discarded weight is $1\times10^{-7}$.  

The ground state (GS) phase diagram, shown in Fig. \ref{diagrama} (a), displays the regions of the above-mentioned phases, illustrated in Fig. \ref{diagrama}(b), 
including the estimated transition lines and the crossover region.
A special feature of the $AB_2$ chain is its symmetry \cite{CINV,SierraPRB1999,*sierraprb2005,MontenegroPRB2006,*OliveiraPRB2009} under the exchange of the
labels of the $B$ sites in a given unit
cell $l$ [identified in the FERRI state, Fig. \ref{diagrama}(b)]. This symmetry implies in a conserved parity $p_l=\pm 1$ in each cell of the lattice. 
The phase diagram of a chain with $N_c$ unit cells is calculated by obtaining the lowest energy for all subspaces with 
$x$ contiguous cells of parity $-1$ and the others $N_c-x$ cells with parity $+1$, with $x=0\ldots N_c$, for fixed $\delta$ and $J$. 
In the phase diagram shown in Fig. \ref{diagrama}(a), $p\equiv\sum_{i=1}^{N_c}p_l =+1$ for $\delta\geq1/3$, $p\neq\pm1$ in the PS region,
and $p=-1$ for $\delta<\delta_{PS,J}$. The magnetic configuration of a phase is identified by the total spin $S_{GS}$, local magnetization, magnetic structure
factor, and spin correlation functions.
In what follows, we shall characterize the phases shown in Fig. \ref{diagrama}(a).
\begin{figure}
\centering{\includegraphics*[width=0.35\textwidth,clip]{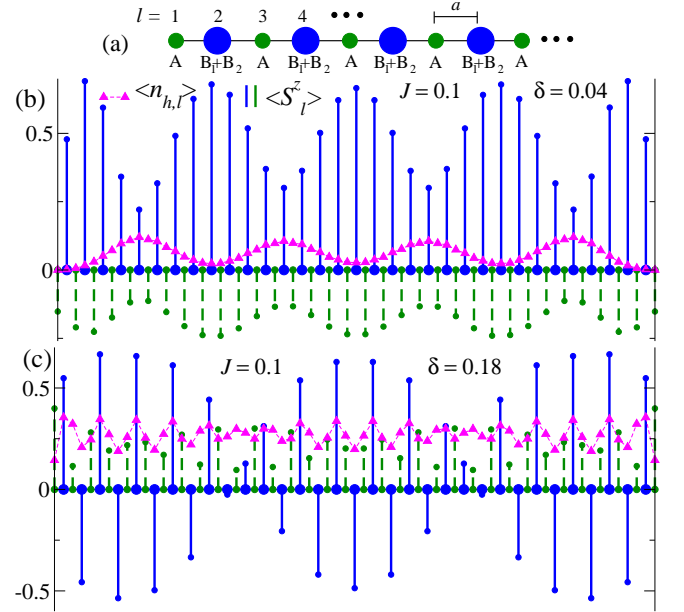}}
\includegraphics*[width=0.48\textwidth,clip]{fig2b.eps}
\includegraphics*[width=0.48\textwidth,clip]{fig2c.eps}
\caption{(Color online). (a) Effective linear chain (spacing $a\equiv 1$) associated with $N=3N_c+1=100$ sites for $J=0.1$ used to illustrate the hole, $\langle n_{h,l}\rangle$, and spin,
$\langle S^z_l\rangle$, profiles: (b) $\delta=4/100$ (FERRI phase) and (c) $\delta=18/100$ (IC phase).\label{densidadesFerri}}
\end{figure}

\section{Ferrimagnetism and transition to IC states}
At $\delta=0$ and $J\neq0$, the \textit{insulating} Lieb ferrimagnetic state with total spin quantum number $S_{GS}=S_{L}\equiv N_c/2-0.5\equiv S_{L}$ is found
for a chain with open boundary conditions, $N=3N_c+1=100$, with an $A$ site on each side. In order to evaluate the stability of this state against doping, we
calculate $S_{GS}$ as a function of $\delta$ from the energy degeneracy in $S^z$. 
As shown in Fig. \ref{diagrama}(c), as hole doping increases from $\delta=0$ to a critical
value $\delta=\delta_{FERRI,J}$, the value of $S_{GS}$ \textit{decreases} linearly
from $S_L$ to 0 or a residual value, signaling a smooth transition to the IC phase. However, for low enough $J$, $S_{GS}$ of the IC phase \textit{increases} linearly
with $\delta$ up to $\delta=\delta_{PS,J}$, the line at which PS occurs [see Fig. \ref{diagrama}(a)], or up to the boundary, $J_{F,\delta}$, of the Nagaoka F phase. 
This unexpected behavior claims for an explanation.
\begin{figure}
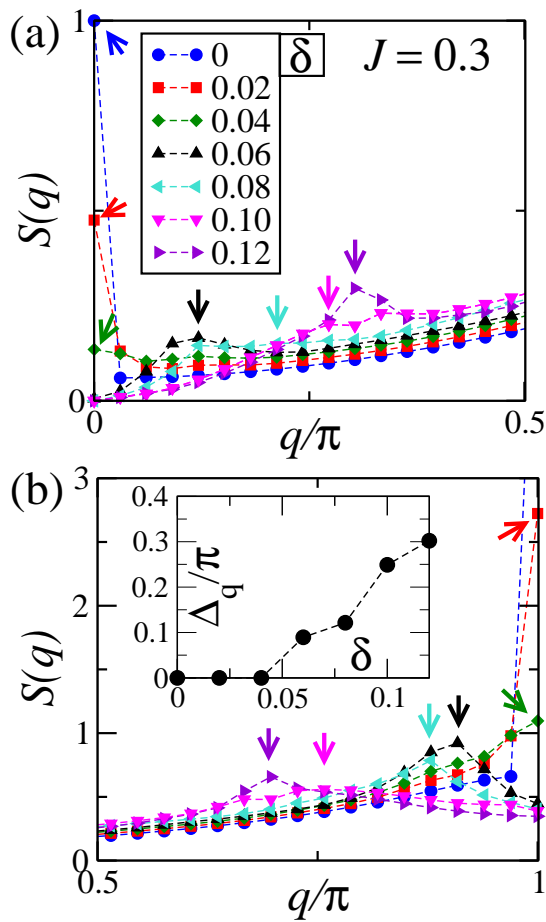

\begin{center}
\includegraphics*[width=0.4\textwidth,clip]{fig3a.eps}
\includegraphics*[width=0.4\textwidth,clip]{fig3b.eps}
\caption{(Color online). Chain with $N=3N_c+1=100$ and $J=0.3$. (a) and (b): Magnetic structure factor $S(q)$ for the indicated 
values of $\delta$. Inset of (b): $\Delta_q\equiv q_{\text{max}}-\pi$, where $q_{\text{max}}$
is the value of $q$ at which the local maximum of $S(q)$, near $q=\pi$, is observed.\label{Fig4}}
\end{center} 
\end{figure}

In order to understand the behavior of $S_{GS}$ for low $J$ we have 
calculated the profiles of the magnetization, $\langle S^z_l\rangle$, in the spin sector $S^z=S_{GS}$, and of the hole density, $\langle n_{h,l}\rangle$, 
for $J=0.1$ (see Fig. \ref{densidadesFerri}).
To help in the data visualization, we use a linearized version of the lattice, as illustrated in Fig. \ref{densidadesFerri}(a).
As shown in Fig. \ref{densidadesFerri}(b), for $\delta = 0.04$ the holes distort the ferrimagnetic structure, which display a modulation  
with wavelength $\lambda \approx 17$, in anti-phase with that exhibited by the hole (charge) density wave. We have thus identified a \textit{modulated itinerant} ferrimagnetic phase
in this underdoped regime. 
On the other hand, as shown in Fig. \ref{densidadesFerri}(c), for $\delta=0.18$ the magnetization has local maxima in coincidence with those of the hole density profile.
In this case, the IC phase is characterized by the presence of \textit{ferromagnetic Nagaoka spin polarons} \cite{PhysRevB.85.245113,polaron} due to hole density wave 
with $\lambda\approx 4$. Our results point 
to a value of $J$ ($\sim 0.2$) below which ferromagnetic ``bubbles'' appear as precursors of the F phase found for $J < J_{F,\delta}$ [see Fig. \ref{diagrama}(a)].

For $J=0.3$, $S_{GS}=0$ in the IC phase, as shown in Fig. \ref{diagrama}(c). 
In Figs. \ref{Fig4} (a) and (b) we present the magnetic structure factor
\begin{equation}
S(q)=\frac{1}{S_{L}(S_{L}+1)}\sum_{l,m}^{2N_c+1}e^{iq(l-m)}\langle\mathbf{S}_l\cdot\mathbf{S}_m\rangle,
\end{equation}
where $l$, $m$ and $\mathbf{S}$ refer to the lattice representation shown in Fig. \ref{densidadesFerri}(a),
for this value of $J$ and doping ranging from $\delta=0$ up to $\delta=0.12$. In a long-range ordered ferrimagnetic state, sharp maxima 
at $q=0$ (ferromagnetism) and $q=\pi$ (antiferromagnetism) are observed in the curve $S(q)$ for $\delta=0$. 
Adding two holes to the undoped state, sharp maxima at $q=0$ and $\pi$ 
are also observed, while broad maxima occur for $\delta=0.04$, indicating short-range ferrimagnetic order which evolves 
to the IC phase by increasing doping, before phase separation (IC-RVB) at the line $\delta=\delta_{PS,J}$ [see Fig. \ref{diagrama}(a)].
In the inset of Fig. \ref{Fig4}(b) we show the departure of the maximum of $S(q)$ from $q=\pi$.

\begin{figure}
\begin{center}
\includegraphics*[width=0.46\textwidth,clip]{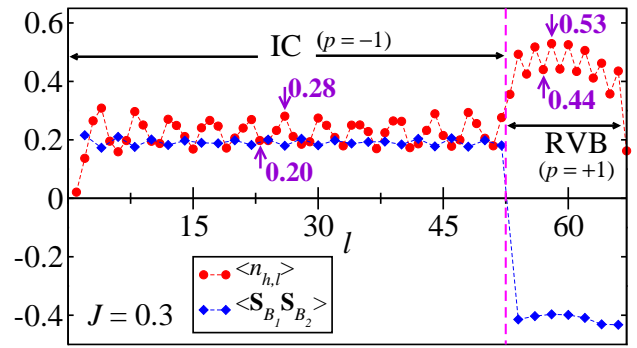}
\caption{(Color online). Phase separation (IC-RVB) for a chain with $N=3N_c+1=100$ sites, $J=0.3$, and $N_h=18$ holes: spin correlation function between $B$ spins at the same cell,
$\langle\mathbf{S}_{B_1,l}\cdot\mathbf{S}_{B_2,l}\rangle$, and 
hole density profile, $\langle n_{h,l}\rangle$.\label{Fig4}}
\end{center} 
\end{figure}

\section{Phase Separation, RVB states and Luttinger Liquid}
In Fig. \ref{diagrama}(a) the dashed line inside the PS region fix the boundary between two 
types of phase separation: in one case, the separation occurs between Nagaoka ferromagnetism and short-range RVB states (F-RVB);
while in the other, it occurs between IC and short-range RVB states (IC-RVB). Indeed, for $0\leq J \lesssim 0.063$ and $\delta_{F-RVB}\leq\delta<1/3$, 
the GS phase separates with F and short range RVB states under coexistence, where $\delta_{F-RVB}$ denotes hole density values along the phase separation line F-RVB,
thereby extending our previous result \cite{MontenegroPRB2006,OliveiraPRB2009} valid only for $J=0$.
However, for $0.063\lesssim J \leq 0.4$ the system behaves differently. The new PS (IC-RVB) region is here illustrated for $J=0.3$, $N=3N_c+1=100$ sites, 
and $N_h=18$ holes:
we thus find that there are 26 cells with odd parity ($p_l=-1$), associated with the IC phase, and the
remaining 7 cells with even parity ($p_l=+1$), associated with the RVB phase. 
In this case, as shown in Fig. \ref{Fig4}(c), the hole-poor IC phase presents a local spin correlation
function $\langle \mathbf{S}_{B_1,l}\cdot \mathbf{S}_{B_2,l}\rangle \approx 0.2$, 
average hole density per site $\approx 0.16$, estimated from the sites indicated by arrows [one $A$ site and two $B$ sites in the 
context of the effective linear chain shown in Fig. \ref{densidadesFerri}(a)], and hole-density wave with $\lambda\approx 4$;
while the hole-rich RVB phase presents $\langle \mathbf{S}_{B_1,l}\cdot \mathbf{S}_{B_2,l}\rangle \approx -0.4$ and 
average hole density per site $\approx 1/3$, estimated from a cell with $A$ and $B$ sites indicated by arrows.
Therefore, apart from boundary effects, the above results thus indicate that the phase separation for a given $J$ value is defined
by the coexistence of the two phases with the hole densities $\delta_{IC\text{-}PS}$ ($\approx 0.16$ for $J=0.3$) and $\delta_{PS\text{-}RVB}$ ($\approx 1/3$ for $J=0.3$) fixed
at the IC-PS and PS-RVB boundaries, respectively, while
the size of the phases are fixed by the chemical doping $\delta=N_h/N$ ($=0.18$ for $N=100$ and $N_h=18$).
We also remark that the stable RVB phase observed at $\delta=1/3$ and $0\leq J\leq 0.4$, which has finite charge and spin gaps, is in agreement with predictions 
for $J=0.35$ \cite{SierraPRB1999,*sierraprb2005} and $J=0$ \cite{MontenegroPRB2006,OliveiraPRB2009}.
\begin{figure}
\begin{center}
\includegraphics*[width=0.48\textwidth,clip]{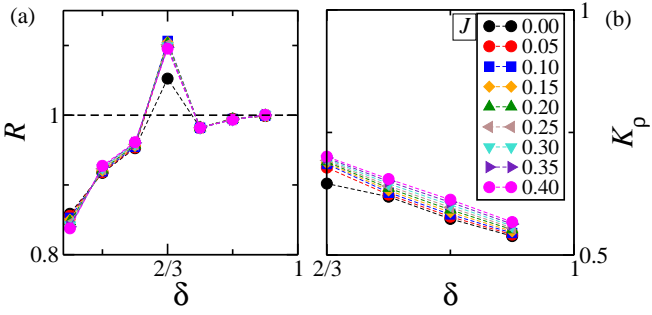}
\caption{(Color online). Luttinger liquid behavior for a chain with $N=3N_c=24$ (ED results). (a) Ratio 
$R=u_{\rho}/\sqrt{D\chi/\pi}$ as a function of $\delta$ for the indicated values of $J$. (b) Exponent $K_{\rho}$ as a
function of $\delta$.\label{Fig6}}
\end{center} 
\end{figure}

For $0 \leq J \leq 0.4$ and $1/3 < \delta < 2/3$, a crossover region with the presence of long-range 
RVB states after hole addition away from $\delta=1/3$ is observed [see Fig. \ref{diagrama}(a)]. At the commensurate filling $\delta=2/3$, the system presents 
a charge gap, while the spin excitation is gapless, also extending our previous result for $J=0$ \cite{MontenegroPRB2006,OliveiraPRB2009}.
\begin{figure}
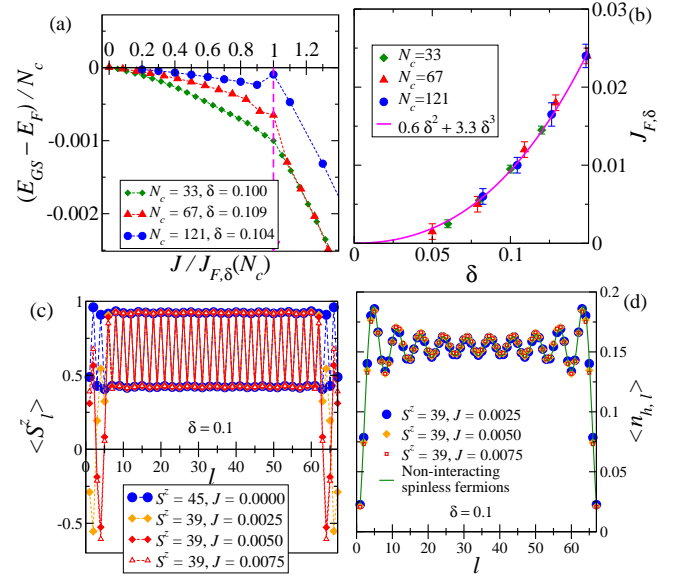

\includegraphics*[width=0.24\textwidth,clip]{fig6a.eps}
\includegraphics*[width=0.23\textwidth,clip]{fig6b.eps}
\includegraphics*[width=0.24\textwidth,clip]{fig6c.eps}
\includegraphics*[width=0.225\textwidth,clip]{fig6d.eps}
\caption{(Color online). (a) Shift $E_{GS}-E_F$ per unit cell, where $E_F$ is the energy of the fully polarized ferromagnetic state, as a function of 
$J/J_{F,\delta}$ for the indicated values of $N_c$ and $\delta$. 
(b) Instability line of the Nagaoka ferromagnetic phase. (c) Spin and (d) hole profiles, $\langle n_{h,l} \rangle$ and 
$\langle S^z_l\rangle$, respectively, for a chain with $N=3N_c+1=100$, $\delta=0.1$, and the indicated values of $S^z$ and $J$.\label{nag}}
\end{figure}

With the aim of investigating the LL behavior as a function of $J$ and $\delta\geq 2/3$, we have 
calculated, through ED techniques, the ratio
$
R=u_\rho/\sqrt{D\chi/\pi},
$
where 
\begin{equation}
\chi=\frac{N_c}{4}[E(N_h+2)+E(N_h-2)-2E_{GS}(N_h)]
\end{equation}
is the charge susceptibility, and $E(N_h\pm2)$ is the total energy for $N_h\pm 2$ holes;
\begin{equation}
D=\frac{N_c}{4\pi}\left[\frac{\partial^2 E(\Phi)}{\partial \Phi^2}\right]_{\Phi_{min}}
\end{equation}
is the Drude weight, where $E(\Phi)$ is the lowest energy for a magnetic flux $\Phi$ through a closed chain, and  
$\Phi_{min}$ its value at $E_{GS}$;
\begin{equation}
u_\rho=\frac{{E(k_{GS}+\Delta k,S=0)-E_{GS}(k_{GS},S_{GS}=0)}}{\Delta k}
\end{equation}
is the charge excitation velocity, where $\Delta k=2\pi/N_c$, and $E(k_{GS}+\Delta k,S=0)$ 
is the lowest energy with wavenumber $k=k_{GS}+\Delta k$ and total spin $S=S_{GS}=0$.
If the low-energy physics of the system is that of 
a LL, we should find $R=1$ \cite{poilblanc}; moreover, the exponent governing the asymptotic behavior of the correlation functions, 
$K_\rho$, satisfies the relation $K_\rho=\pi u_\rho/2\chi$. As shown in Fig. \ref{Fig6}(a), $R$ is indeed very close 
to 1 for $\delta>2/3$; 
in addition, as shown in Fig. \ref{Fig6}(b), we find $0.7 \gtrsim K_\rho \gtrsim 0.5$ for $\delta>2/3$. 
Remarkably, as shown in Figs. \ref{Fig6}(a) and (b), the data for $R$ and $K_\rho$ exhibit data collapse 
as a function of $\delta$ for $0 \leq J \leq 0.4$.
In short, the results above clearly indicate that for $\delta>2/3$ and $0\leq J\leq 0.4$ the 
system behaves as a LL in the strong coupling regime.

\section{Stability of Nagaoka ferromagnetism}

In this Section, we shall provide strong evidence that
for $0\leq J\leq J_{F,\delta}$ and $0<\delta \leq \delta_{F-RVB}$, the kinetic energy of holes is
lowered in a fully polarized ferromagnetic state, an extension of Nagaoka ferromagnetism \cite{nagaoka,tasaki},
with the GS energy equal to that of non-interacting spinless fermions: $E_{GS}=E_F$.

The estimate of $J_{F,\delta}$ is based on the data for the shift $(E_{GS}-E_F)/N_c$ as a function of $J$, as illustrated in Fig. \ref{nag}(a)
for $\delta$ close or equal to 0.1.
We stress that the shift decreases as $N_c$ increases for $0 < (J/J_{F,\delta}) < 1$, and goes to zero in the thermodynamic
limit. In addition, one should notice that, by examining the data above and below $J=J_{F,\delta}$, particularly for $N=3N_c+1=364$ sites, 
$\partial{E_{GS}}/\partial{J}$ appears to be discontinuous at $J=J_{F,\delta}$ in the thermodynamic limit, thus suggesting a first-order transition
to the IC phase at $(J/J_{F,\delta})=1$. In Fig. \ref{nag}(b) we show that our estimated transition line, $J_{F,\delta}$, [see also Fig. \ref{diagrama}(a)] 
is almost not affected by finite size effects and implies $\delta_{F,J}\sim \sqrt{J}$ as $\delta\rightarrow 0$, 
as found from analytical results \cite{Altshuler,PhysRevB.85.245113} for the $t$-$J$ model in a square lattice.
In particular, for $J=0$ the instability of the Nagaoka state occurs at $\delta\approx 0.23$, which is very close to the values of hole doping estimated
for $n$-leg ladder systems \cite{PhysRevLett.108.126406} and the square lattice \cite{PhysRevLett.108.126406,2D_F}.

The spin profile for a chain with $N=3N_c+1=100$, $J\gtrsim 0$ and $\delta=0.1$ is also in very good agreement with the Nagaoka state, as shown in Fig. \ref{nag}(c), although 
boundary effects are visible for $J\gtrsim 0$; in fact, $S^z$ changes from 45 to 39 (on average, three spins at each boundary are 
not fully polarized), but one should notice that the change saturates as $J$ slightly 
increases above zero. This fact is corroborated by the hole density shown in Fig. \ref{nag}(d), whose data 
for the referred states with $S^z=39$ at $\delta=0.1$ are very well described by the Nagaoka profile.

\section{Discussion and Concluding Remarks} 

The presented phase diagram of doped $AB_2$ $t$-$J$  chains is remarkably rich. Indeed,
several magnetic and nonmagnetic phases manifest themselves in a succession of surprising relevant features, some of which are
similar to those observed in the square and $n$-leg ladder lattices: all in a simple doped chain. 
In particular, we emphasize 
the modulated ferrimagnetic structure, the occurrence of Nagaoka spin polarons in the underdoped regime and small values of $J$, 
incommensurate structures with nonzero magnetization, 
the strong-coupling LL physics in the high-doped regime, and
the instability of Nagaoka ferromagnetism against $J$ and doping.
Therefore, these chains are unique systems and of relevance for the
physics of polymeric materials, whose properties might also represent challenging topics to be explored via analog simulations in
ultracold fermionic optical lattices.

This work was supported by CNPq and FACEPE through the PRONEX program, and CAPES (Brazilian agencies).
\bibliography{tjbib}
\end{document}